\documentclass{icrc29}
\usepackage{graphicx,amssymb,amsmath,times}
\begin{document}
%Title of paper
\title[Search for Cygnus Arm Diffuse ...]{Search for Cygnus Arm Diffuse TeV Gamma Ray Emission with the
Whipple 10m Telescope}
\author[R. Atkins et al] {R. Atkins$^a$ for the VERITAS Collaboration$^b$\\
        (a) High Energy Astrophysics Institute, University of Utah, Salt Lake City, Utah, USA\\
        (b) For full author list, see J. Holder's paper ``Status and Performance of the First VERITAS Telescopes'' from these proceedings
      }
\presenter{Presenter: R. Atkins (ratkins@cosmic.utah.edu), \  
usa-atkins-R-abs1-og21-poster}

\maketitle

\begin{abstract}
The Whipple 10 meter atmospheric Cherenkov telescope has 
made observations of the region known as the Cygnus arm.  
This region has been recently reported by the Milagro 
experiment to contain a diffuse TeV $\gamma$-ray source 
centered at RA=308 and Dec=42.  We report upper limits 
(using the Whipple 10 m telescope) obtained during the Fall 
2004 observing season centered on RA=310 and Dec=42.65. 

\end{abstract}

\section{Introduction}
The Milagro observatory\cite{milagro_all_sky} has made long term observations 
of the Cygnus Arm.  They report an excess of over 5.5$\sigma$ over 
a 5.9$^{\circ}$ square bin in RA and Dec.\cite{dr_smith}.  This excess is 
inconsistent with a point source and may be due to a giant molecular 
cloud(GMC) located in the same region as the excess. This cloud has been 
reported by Dame et. al.\cite{dame_2001,dame_1985} to be at a distance of 
1.7 pc with a estimated mass of $5.1\times 10^{6}M_{\odot}$.  The angular 
extent of the cloud is 44 square degrees.  

Diffuse emission of $\gamma$ rays at TeV energies have long been speculated 
to be the result of cosmic ray interactions with giant molecular 
clouds\cite{Gould_1965,pollack_1963}.  In this scenario, galactic cosmic 
rays interact with hydrogen and produce neutral pions.  These pions quickly 
decay and produce $\gamma$ rays.  Predictions by Aharonian and 
Atoyan \cite{Aharonian_2001,Aharonian_1996} have indicated that the flux 
from these GMC should follow the galactic cosmic ray flux (excluding 
enhancements by local sources) and would be proportional to the GMC Mass over the 
square of the distance to the GMC.  The CygX cloud is a good target since it is close and 
very massive.

\section{Analysis}
The Whipple 10 meter atmospheric Cherenkov telescope utilizes the well 
proven imaging technique to reject cosmic ray background events and to 
determine source geometry\cite{Weekes_1989,Mohanty_et_al}.  This method 
uses the shape of the shower image (fitted to an ellipse) to determine 
if the shower was initiated by a $\gamma$ primary 
or a cosmic ray primary.  Additionally, if the source is assumed to be at 
the center of the field of view (FOV), the angle between the 
major axis of the ellipse and the line formed by the centroid of the 
image and the center of the FOV($\alpha$ angle), can be used to eliminate 
events not coming from the source location.  The energy threshold for the Whipple 
10 meter is 390 GeV for a Crab like spectrum\cite{finley}

Extensions of this method have been made to make observations for objects 
that may not be in the center of the FOV.  This is often the case when 
searching for new sources, diffuse emission, or sources that have been 
identified by other experiments with relatively low angular resolution. 
In this two dimensional analysis \cite{Lessard_2001}, the source location 
is geometrically constrained to lie along the major axis of the shower image 
(as it the case with the one dimensional analysis), but no 
requirement is made of the $\alpha$ angle with respect to the center 
of the camera.  The distance from the image centroid to the source location along the major axis 
is estimated using  
\begin{equation}
d = \xi\left( 1 - \frac{width}{length}\right )
\end{equation}
where the {\it width} refers to the size of the minor axis, {\it length} refers to 
the size of the major axis, {\it d} is the distance along the major axis, and $\xi$ 
is a scaling parameter that must be determined.  To break the ambiguity as to which 
direction along the major axis the source lies, the skewness in the image is used.

The $\xi$ parameter was determined by examining the crab supernova 
remnant \cite{Weekes_1989,1991ApJ...377..467V}. The two dimensional 
analysis was applied to on-source crab data.  To optimize the $\xi$ parameter,
the value of $\xi$ was varied in steps of $\delta \xi = 0.05^{\circ}$.  The 
optimal value was determined by the maximum signal at the source location   
The optimal value was determined to be $\xi=1.35^{\circ}$. 
  
%\begin{figure}[h]
%\begin{center}
%\includegraphics*[width=0.7\textwidth, height = 0.2\textheight,angle=0,clip]{fig1.eps}
%\caption{\label {fig1} Excess number of counts as a function of scale parameter 
% $\xi$.  In general the optimal value will depend upon the zenith angle of the observations and the 
%energy.  However, for small zenith angles (Z $<$ 35$^{\circ}$) these effects can be neglected}  
%\end{center}
%\end{figure} 

Once the $\xi$ parameter has been determined the data can binned and the point spread function (PSF) for the 
method can be determined.  Here we have used a 0.36$^{\circ} \times 0.36^{\circ}$ square bin in RA and Dec.  This bin 
size was found to optimize the significance of the on source crab observations.  The binning of the 
data is shifted six times in RA and Dec. in steps of 0.06$^{\circ}$ in order to compensate for edge effects 
in the binning.  Applying this analysis to the on source Crab data we get a maximum significance 
of 11.6$\sigma$ from 5.6 hours of on source data (4.9$\sigma$/$\sqrt{hr}$).  The PSF of the excess in 
RA and Dec. is fit to a Gaussian distribution with a $\sigma_{PSF}$ = 0.18$^{\circ}$

For points source off axis (that is to say, not in the center of the field) the PSF becomes broader as the 
source moves further away from the center of the FOV.   While the radial spread to the PSF stays roughly the same, 
the azimuthal spread increases slightly from 0.18$^{\circ}$ to 0.21$^{\circ}$ at one degree offset.  The behavior of the PSF as 
function off offset was determined by analyzing crab data taken at 0.3, 0.5, 0.8 and 1.0 degree offsets from 
the center of the field.
 
\begin{figure}[h]
\begin{center}
\includegraphics*[width=.9\textwidth,height = 0.23\textheight,angle=0,clip]{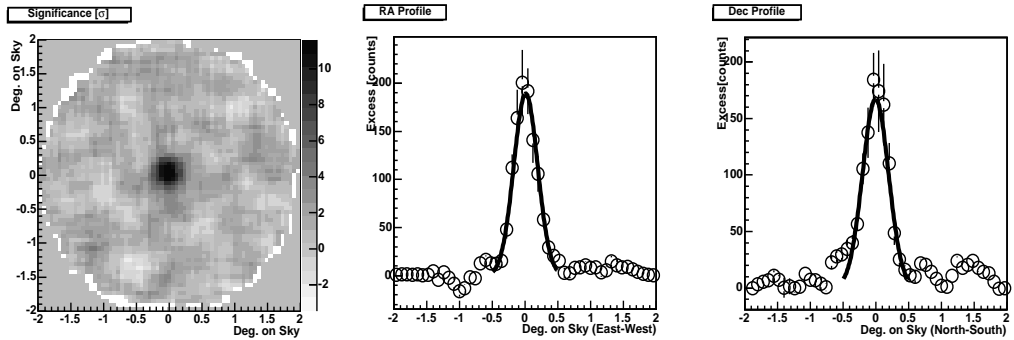}
\caption{\label {fig2} Application of the two dimensional analysis to the Crab supernova remnant.  The maximum 
significance (pre trials) is 11.6 $\sigma$.  The profiles in degrees on sky show a good fit to the center 
width of $\sigma_{PSF}$ = 0.18$^{\circ}$. and to the location of the source  }  
\end{center}
\end{figure} 

\section{Data}
Data used in this work was taken during the months of August 2004 through November 2004.  The 
observation window for this object is small as the Whipple 10 meter generally suspends 
observations in the summer months due to poor weather conditions in southern Arizona.  In this analysis 
we have used 12 ON/OFF pairs of 28 minutes each.  The total number of events in the ON/OFF field after shape cuts 
is 14406/14594 (ON/OFF).  The coordinates of the observations are 
RA = 20:40:7.9 (310.03$^{\circ}$) and Dec = 42:39:51.12 (42.66$^{\circ}$) in J2000 coordinates.
These coordinates were chosen to overlap with the morphology of the Milagro 
excess \cite{dr_smith} as well as overlap with large values of 
neutral hydrogen column densities in the region \cite{dame_2001}.

\section{Results and Discussion}
The above analysis fails to find strong evidence for a point source of $\gamma$-rays within the 
2-D FOV of the observations.  Figure 2 shows the excess map and sigma map from the field.  
The significance was calculated using the standard Li and Ma method \cite{li_ma}.
\begin{figure}[h]
\begin{center}
\includegraphics*[width=0.85\textwidth,height = 0.3\textheight,angle=0,clip]{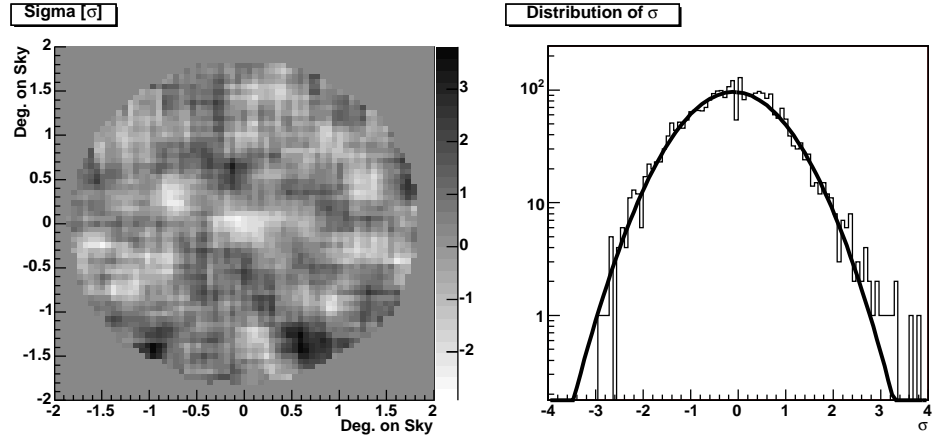}
\caption{\label {fig3} (Left) Sigma map of the Cygnus Arm FOV in degrees on sky. relative to the center of 
the field (RA = 20:40:7.9 (310.03$^{\circ}$) and Dec = 42:39:51.12 (42.66$^{\circ}$)) (where north is 
up and east is to the right).  (Right) Distribution of significance.  The most significant 
bin (located at RA = 310.8$^{\circ}$, Dec = 41.3$^{\circ}$) has a pre-trials significance of 3.8$\sigma$ with a 
post-trial probability of being a chance fluctuation of 12\%}
\end{center}
\end{figure}
The most significant bin in the map (Figure 2) is located at RA=310.8$^{\circ}$ and Dec=41.3$^{\circ}$.  The pretrial significance is 
3.8$\sigma$ in this bin.  To account 
for trials factors associated with the binning and the oversampling we simulated 30,000 data sets for 
this field.  We find the chance probability of getting one bin with a sigma of 3.8 or higher is 12\%  

As no compelling point source was found within this field of view,
we must conclude that the Milagro source\cite{dr_smith} must be rather diffuse in
this region, or must be at a flux level below the sensitivity of this analysis.  
Diffuse sources are rather difficult with the Whipple 10 meter, particularly if 
size of the regions of emission is extends beyond the FOV.  Such emission would be difficult to detect 
with this technique, which is optimized for point-source detection.

Upper limits have been calculated in the manner prescribed by Helene\cite{Helene}.  Once the upper limit is known, the 
values are adjusted to correct for the changing sensitivity across the FOV.  This changing sensitivity has been 
determined by the analysis of offset Crab data.  Figure 3 shows the upper limits for each bin in RA and Dec. The field has 
been limited to the inner 1$^{\circ}$ of the FOV.  Beyond this the sensitivity decreases such that it is difficult 
to set meaningful upper limits. 
\begin{figure}[h]
\begin{center}
\includegraphics*[width=1.0\textwidth,height = 0.25\textheight,angle=0,clip]{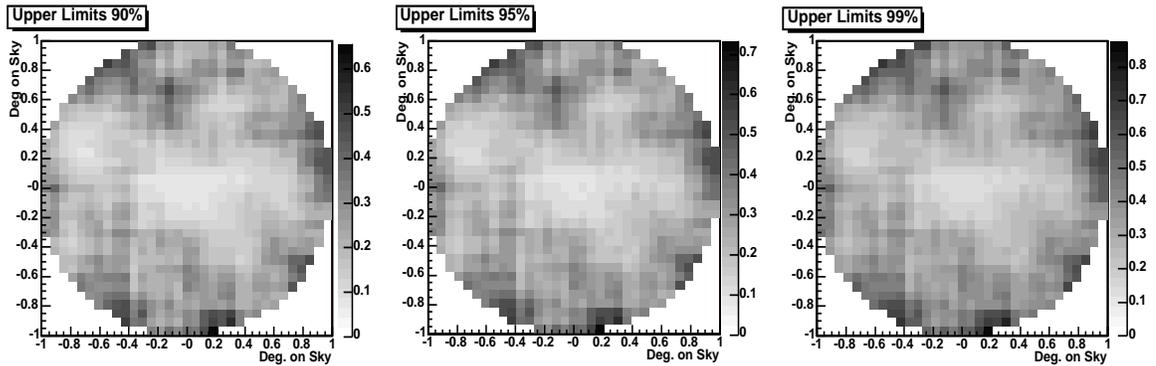}
\caption{\label {fig5}90\%, 95\%, and 99\% upper limits, in $\gamma$ /minute for the inner 1$^{\circ}$ FOV   
of the field.}
\end{center}
\end{figure}
We find no point source within in the inner half degree of the 
FOV above 3.0\%, 3.3\%, and 4.0\% of the Crab at the 90\%, 95\% and 99\% 
confidence level.  In the outer half degree we find 
no point source above 15.0\%,16.7\% and 20\% of the Crab at the 90\%, 95\%, 
99\% confidence level.    These upper limits are obtained by comparing the 
total number of excess counts at a given offset (within the 2$\sigma$ ellipse of the 
PSF) with the expected excess counts from the Crab at the same offset.  
This method assumes a crab like spectrum and point source distribution. 

Further work is being done to convolute the PSF with the expected source distribution, 
obtained from neutral hydrogen column density maps of the region. This will allow us to set 
stronger upper limits for the region and may reveal a sources that the point source analysis failed to detect.  
We are also continuing to collect data on this target and are also 
observing other fields within the region of the Milagro excess.

\section{Acknowledgments}
This work supported by the US National
Science Foundation Grant (\#PHY 0099580), the U.S. Department of Energy, the Smithsonian Institute, the NSERC in Canada, the 
Science Foundation of Ireland and by PPARC in the United Kingdom.


\begin{thebibliography}{99}

\bibitem{Aharonian_2001} Aharonian, F.~A.\ 2001, 
Space Science Reviews, 99, 187 

\bibitem{Aharonian_1996} Aharonian, F.~A. \& 
Atoyan, A. M. 1996, Astron. \& Astro. , 309, 917

\bibitem{milagro_all_sky} 
Atkins, R., et al., 2004, ApJ. , 608, 680.

\bibitem{dame_2001} Dame, T.~M., Hartmann, D., 
\& Thaddeus, P.\ 2001, ApJ, 547, 792. 

\bibitem{dame_1985} Dame, T.~M., \& 
Thaddeus, P.\ 1985, ApJ, 297, 751 

\bibitem{finley} Finley, J.~P., \& The VERITAS Collaboration 2001, Proceedings of the 27th 
International Cosmic Ray Conference.~07-15 August, 2001.~Hamburg, 
Germany, p.2827, 27, 2827.
 
\bibitem{Gould_1965} Gould, R.~J.~\& 
Burbidge, G.~R.\ 1965, Annales d'Astrophysique, 28, 171 

 
\bibitem{Helene} Helene, O., Nucl.Inst.\& Meth. A212 (1983) 319-322.

\bibitem{Lessard_2001} Lessard, R.~W., 
Buckley, J.~H., Connaughton, V., \& Le Bohec, S.\ 2001, Astroparticle 
Physics, 15, 1 

\bibitem{li_ma} Li, T.-P., \& Ma, Y.-Q.\ 
1983, ApJ, 272, 317. 

\bibitem{Mohanty_et_al} Mohanty, G., et al.\ 
1998, Astroparticle Physics, 9, 15  

\bibitem{pollack_1963} Pollack, J.~B.~\& 
Fazio, G.~G.\ 1963, Physical Review , 131, 2684

\bibitem{dr_smith} 
Smith, A. et al., 2004, Heidelberg TeV Gamma Ray Workshop, in proceedings.

\bibitem{1991ApJ...377..467V} Vacanti, G., et al.\ 
1991, ApJ, 377, 467 

\bibitem{Weekes_1989} Weekes, T.~C., et al.\ 
1989, ApJ, 342, 379

\end{thebibliography}
\end{document}